\documentclass{elsart}

\usepackage{epsfig}% Include figure files
\newcommand{\mpp}{M_{p\bar{p}}}
\newcommand{\mb}{{M_{\rm bc}}}
\newcommand{\de}{{\Delta{E}}}
\newcommand{\bp}{{B^{+}}}

\newcommand{\pp}{{p\bar{p}}}
\newcommand{\ppk}{{p\bar{p}K^+}}

\newcommand{\ppks}{{p\bar{p}K_S^0}}

\newcommand{\plpi}{{p\bar{\Lambda}\pi^-}}

\newcommand{\ks}{{K_S^0}}

\begin{document}

\begin{flushright}
                  \noindent \hspace*{3.0in}{
                   % Belle Note 711 v0.4
%                    BELLE-CONF-0413
%                  \\ICHEP Abstract 11-0658
                    Belle Preprint 2005-13 \\
                    KEK \ Preprint 2005-4
}
\end{flushright}

%\preprint{\vbox{ \hbox{   }
%                  \hbox{draft V0}
%                  \hbox{Belle Note 712}
%                  \hbox{BELLE-CONF-0423}
%                  \hbox{ICHEP Abstract 11-0670}
%                 \hbox{EPS-ID 531}
%                 \hbox{hep-ex/0302024}
%}}
\begin{frontmatter}
\epsfysize3cm
\vskip -10mm
\title{ \quad\\[1cm]\large
Study of the Baryon-Antibaryon Low-Mass Enhancements in 
Charmless Three-body 
Baryonic $B$ Decays}

%% >>>> Insert final authorlist here

%%% Paper:    B -> p pbar h(*)
%%% Journal:  Physical Review Letters
%%% Contacts: M. Wang (mwang@phys.ntu.edu.tw)
%%% Paper:    B -> p p-bar K, p Lambda-bar pi
%%% Journal:  Physics Letters B
%%% Contacts: M. Wang (mwang@phys.ntu.edu.tw)
%%%           T.-L. Kuo (cathy@hep1.phys.ntu.edu.tw)
%%% Non-responding authors or those who said NO are commented out.
%%% ====================================================================
%%% Click the RELOAD button on your web browser to see the updated file.
%%% ====================================================================
%%% Use \input{author} to insert this material into your latex file.

\collab{Belle Collaboration}
   \author[Taiwan]{M.-Z.~Wang}, % Taiwan
   \author[KEK]{K.~Abe}, % KEK
   \author[TohokuGakuin]{K.~Abe}, % TohokuGakuin
% \author[TIT]{N.~Abe}, % TIT
% \author[KEK]{I.~Adachi}, % KEK
   \author[Tokyo]{H.~Aihara}, % Tokyo
% \author[Nagoya]{M.~Akatsu}, % Nagoya
   \author[Tsukuba]{Y.~Asano}, % Tsukuba
% \author[Toyama]{T.~Aso}, % Toyama
   \author[BINP]{V.~Aulchenko}, % BINP
 \author[ITEP]{T.~Aushev}, % ITEP
% \author[Tata]{T.~Aziz}, % Tata
   \author[Cincinnati]{S.~Bahinipati}, % Cincinnati
   \author[Sydney]{A.~M.~Bakich}, % Sydney
% \author[ITEP]{V.~Balagura}, % ITEP
% \author[Peking]{Y.~Ban}, % Peking
% \author[Tata]{S.~Banerjee}, % Tata
% \author[Melbourne]{E.~Barberio}, % Melbourne
% \author[Hawaii]{M.~Barbero}, % Hawaii
% \author[Lausanne]{A.~Bay}, % Lausanne
   \author[BINP]{I.~Bedny}, % BINP
% ...
% Ljubljana
and
% \author[Lausanne]{D.~Z\"urcher}, % Lausanne

%%%\address[Aomori]{Aomori University, Aomori, Japan}
\address[BINP]{Budker Institute of Nuclear Physics, Novosibirsk, Russia}
% ...
\address[Cincinnati]{University of Cincinnati, Cincinnati, OH, USA}
%%%\address[Frankfurt]{University of Frankfurt, Frankfurt, Germany}
% ...
\address[KEK]{High Energy Accelerator Research Organization (KEK), Tsukuba, Japan}
% ...
\address[ITEP]{Institute for Theoretical and Experimental Physics, Moscow, Russia}
% ...
\address[Taiwan]{Department of Physics, National Taiwan University, Taipei, Taiwan}
% ...
\address[Sydney]{University of Sydney, Sydney, NSW, Australia}
% ...
\address[TohokuGakuin]{Tohoku Gakuin University, Tagajo, Japan}
% ...
\address[Tokyo]{Department of Physics, University of Tokyo, Tokyo, Japan}
% ...
\address[Tsukuba]{University of Tsukuba, Tsukuba, Japan}
%%%\address[Utkal]{Utkal University, Bhubaneswer, India}
% ...
\thanks[NovaGorica]{on leave from Nova Gorica Polytechnic, Nova Gorica, Slovenia}

%\collaboration{Belle Collaboration}
%\noaffiliation
\normalsize

\begin{abstract}
The angular distributions 
of the baryon-antibaryon low-mass enhancements seen in
the charmless three-body baryonic $B$ decays
$B^+ \to p \bar{p} K^+$,
$B^0 \to p \bar{p} K_S^0$, and $B^0 \to p \bar{\Lambda} \pi^-$ are reported. 
A quark fragmentation interpretation
is supported, while the gluonic resonance picture is disfavored.
Searches for the
$\Theta^+$ and $\Theta^{++}$ pentaquarks in the relevant decay modes and
possible glueball states $\mathcal{G}$ 
 with 2.2 GeV/$c^2  <  \mpp < 2.4$ GeV/$c^2$ in the 
$p \bar{p}$ systems
give null results.  We set upper limits on the 
products of branching fractions,  
${\mathcal B}(B^0 \to {\Theta^+}\bar{p})\times {\mathcal B}({\Theta^+
}\to p\ks) < 2.3 \times 10^{-7}$,
${\mathcal B}(\bp \to {\Theta^{++}}\bar{p})\times {\mathcal B}({\Theta^{++}
}\to p K^+) < 9.1 \times 10^{-8}$, and 
${\mathcal B}(\bp \to \mathcal{G} K^+)\times {\mathcal B}(\mathcal{G} \to \pp) 
< 4.1 \times 10^{-7}$ at the 90\% confidence level.
The analysis is based on a
140~fb$^{-1}$ data sample recorded on the $\Upsilon({\rm 4S})$
resonance with the Belle detector at the KEKB asymmetric-energy $e^+e^-$
collider. 

\noindent{\it PACS:} 13.25.Hw, 13.60.Rj
%\pacs{13.25.Hw, 13.60.Rj}
\end{abstract}
\end{frontmatter}
%\clearpage

%\tighten
{\renewcommand{\thefootnote}{\fnsymbol{footnote}}
\setcounter{footnote}{0}

Observations of several baryonic three-body $B$ decays have been
reported recently~\cite{ppk,plpi,pph,LLK}.
One common feature of these observations is the peaking of
the baryon-antibaryon pair mass 
spectra toward threshold, as originally conjectured 
in Refs.~\cite{HS,rhopn} and elaborated more recently in
Refs.~\cite{glueball,RosnerB,rus}. The same peaking behavior
near threshold has been found in baryonic $J/\psi$ decays~\cite{BES} as well,
indicating that this may be a universal phenomenon.
%%% "universal" sounds too strong for only two examples -LEP %%%
Possible explanations include intermediate
(gluonic) resonant states or non-perturbative QCD
 effects of the quark fragmentation process~\cite{glueball,RosnerB}.
Alternatively, the dynamical picture can be replaced
by an effective range analysis with a baryon form factor~\cite{rus}. 
To distinguish among the above hypotheses for the production
mechanism, we study the
angular distributions of the threshold enhancements
in the helicity frame for the
$B^+ \to \ppk$, $B^0 \to \ppks$ and $B^0 \to \plpi$~\cite{conjugate} modes.  
%Also, we update the mass
%spectra from our previous studies. 
%with higher statistics. 

We use a  140 fb$^{-1}$  data sample,
consisting of 152 $ \times 10^6 B\bar{B}$ pairs,
collected with the Belle detector %on the $\Upsilon({\rm 4S})$ resonance
at the KEKB asymmetric energy $e^+e^-$ (3.5 on 8~GeV) collider~\cite{KEKB}.
The Belle detector is a large solid angle magnetic spectrometer
that consists of a three layer silicon vertex detector (SVD), a 50
layer central drift chamber (CDC), an array of aerogel threshold
\v{C}erenkov counters (ACC), a barrel-like arrangement of time of
flight scintillation counters (TOF), and an electromagnetic
calorimeter comprised of CsI(Tl) crystals located inside a
superconducting solenoid coil that provides a 1.5~T magnetic
field.  An iron flux return located outside of the coil is
instrumented to detect $K_L^0$ mesons and to identify muons. The
detector is described in detail elsewhere~\cite{Belle}.

The event selection criteria are based on the information obtained
from the tracking system
(SVD and CDC) and the hadron identification system (CDC, ACC, and TOF).
All primary charged tracks
are required to satisfy track quality criteria
based on the track impact parameters relative to the
interaction point (IP). %, which is determined run-by-run.
The deviations from the IP position are required to be within
$\pm$1 cm in the transverse ($x$--$y$) plane, and within $\pm$3 cm
in the $z$ direction, where the $z$ axis is opposite the
positron beam direction. For each track, the likelihood values $L_p$,
$L_K$, and $L_\pi$ that it is a proton, kaon, or pion, respectively,
are determined from the information provided by
the hadron identification system.  The track is identified as a proton
if $L_p/(L_p+L_K)> 0.6 $ and $L_p/(L_p+L_{\pi})> 0.6$, or as a kaon if
$L_K/(L_K+L_{\pi})> 0.6$, or as a pion if $L_{\pi}/(L_K+L_{\pi})> 0.6$.
%based on
%using $p/K/\pi$ likelihood functions obtained from the hadron
%identification system. For protons, we require $L_p/(L_p+L_K)> 0.6 $ and
%$L_p/(L_p+L_{\pi})> 0.6$, where $L_{p/K/\pi}$ stands for the
%proton/kaon/pion likelihood.
The proton selection efficiency is about 84\% (88\% for $p$ and 80\% for 
$\bar{p}$)
for particles with
momenta at 2 GeV/$c$,
and the fake rate is about
10\% for kaons and 3\% for pions.
%To identify charged kaons (pions), we require the kaon (pion)
%$K$--$\pi$ likelihood ratio to be greater than 0.6. 
%We require $L_K/(L_K+L_{\pi})> 0.6$ to identify kaons and  
%$L_{\pi}/(L_K+L_{\pi})> 0.6$ for pions.
Candidate $\ks$ mesons
are reconstructed from pairs of oppositely charged tracks (both treated as
pions)
%%% via the $\pi^+\pi^-$ decay channel
having a mass consistent with  the $\ks$ nominal mass, 
%%% No pion-ID requirement [as defined above] on these tracks, right? %%%
$ |M_{\pi^+\pi^-} - M_{K^0}| < 30$ MeV/$c^2$,
%The candidate must also 
as well as
a displaced vertex and flight direction consistent with
%%%a $\ks$ originating from the interaction point. 
an origin at the IP.
%with the method described in Ref.~\cite{kpi}.  
%Note that only 
%kinematical information is used for the $\ks$ selection and there is no
%likelihood requirement on the
%secondary pions. 
Candidate $\Lambda$ baryons are reconstructed from pairs of oppositely
charged tracks---treated as a proton and negative pion---whose  mass
is consistent with the nominal $\Lambda$ baryon mass, 
1.111 GeV/$c^2 < M_{p\pi^-} <1.121 $ GeV/$c^2$.  The proton-like
daughter is required to satisfy $L_p/(L_p+L_{\pi})> 0.6$.
%via the $p\pi^-$ decay channel 
%following similar selection procedure for $\ks$. To reduce background, 
%a $L_p/(L_p+L_{\pi})> 0.6$ requirement is applied 
%to the secondary proton from $\Lambda$ decay. 

%Since the initial collision 
%center of mass energy is set to match the $\Upsilon({\rm
%4S})$ resonance, which decays into a $B\bar{B}$ pair, one can use
Candidate $B$ mesons are reconstructed in the $B^+ \to p \bar{p} K^+$,
$B^0 \to p \bar{p} K_S^0$, and $B^0 \to p \bar{\Lambda} \pi^-$  modes.
%an identified proton, an
%identified pion, and a reconstructed $\Lambda$ baryon.
We use two kinematic variables in the center of mass (CM) frame to identify the
reconstructed $B$ meson candidates: the beam energy
constrained mass $\mb = \sqrt{E^2_{\rm beam}-p^2_B}$, and the
energy difference $\de = E_B - E_{\rm beam}$, where $E_{\rm
beam}$ is the beam energy, and $p_B$ and $E_B$ are the momentum and
energy, respectively, of the reconstructed $B$ meson.
%, all in the rest frame of
%the $\Upsilon({\rm 4S})$.k
The candidate region is
defined as 5.20 GeV/$c^2 < \mb < 5.29$ GeV/$c^2$ and $-0.1$ GeV $ < \de< 0.2$
GeV. From a GEANT~\cite{geant} based Monte Carlo (MC) simulation, the signal
peaks in the 
subregion 5.27 GeV/$c^2 < \mb < 5.29$ GeV/$c^2$ and $|\de|< 0.05$ GeV.
The lower bound of $\de$ is chosen to exclude possible contamination from
so-called ``cross-feed'' baryonic $B$ decays.

The background in the fit region arises solely from the continuum $e^+e^-
\to q\bar{q}$ ($q = u,\ d,\ s,\ c$) process.
% Owing to the $\de > $ -0.1 GeV  selection, 
%the  contributions from ``cross-feed'', where similar 
%types of rare decay events pass each other's signal criteria, is negligible.
%Except for a small 
%feed-across between these rare decay modes,
%the dominant background
%is from the continuum $e^+e^- \to q\bar{q}$ process.
%The background from $b \to c$ and charmless mesonic decays is 
%also negligible.
%This is confirmed using
%an off-resonance data set (8.8 fb$^{-1}$)
%taken 60 MeV
%below the $\Upsilon({\rm 4S})$ and a MC sample of 60 million continuum events.
We suppress the jet-like continuum background events relative to the more
spherical $B\bar{B}$ signal events using a Fisher discriminant~\cite{fisher}
that combines seven event shape variables, as described in Ref.~\cite{etapk}.
%In the $\Upsilon({\rm 4S})$ rest frame,
%continuum events are jet-like while
%$B\bar{B}$ events are more spherical. 
%One can use the reconstructed momenta 
%of final state particles to form various shape variables (e.g. thrust
%angle, Fox-Wolfram moments, etc.) in order to categorize each event.  
%We follow the scheme defined in Ref.~\cite{etapk} that
%combines seven event shape variables into
%a Fisher discriminant~\cite{fisher} in order to suppress
%continuum background. 
%The variables chosen have
%almost no correlation
%with $\mb$ and $\de$.
Probability density functions (PDFs) for the Fisher discriminant and
the cosine of the angle between the $B$ flight direction
and the beam direction in the $\Upsilon({\rm 4S})$ rest frame
are combined to form the signal (background)
likelihood ${\mathcal L}_{s}$ (${\mathcal L}_{b}$).
The signal PDFs are determined using signal MC
simulation; the background PDFs are obtained from 
the side-band data %the continuum MC
%simulation for events with
with $\mb < 5.26$ GeV/$c^2$.
We require
the likelihood ratio ${\mathcal R} = {\mathcal L}_s/({\mathcal L}_s+{\mathcal L}_b)$ 
to be greater than 0.7, 0.75, and 0.8 for the
$\ppk$, $\ppks$, and $\plpi$ modes, respectively.
These selection
criteria are determined by optimization of $n_s/\sqrt{n_s+n_b}$, where $n_s$ 
and $n_b$
denote the expected numbers of signal and background events, respectively. 
%Note that a nominal signal branching fraction~\cite{ppk,plpi,pph} is assumed
%for each mode to determine $N_s$.
We use the branching fractions from our 
previous measurements~\cite{plpi,pph} in the calculation of $n_s$.
%suppressing about 94\% of the background while retaining 66\% of the signal.
%In this study, there is only one $B$ candidate allowed per event. 
If there are  multiple $B$ candidates in a single event, we 
select the one with the best $\chi^2$ value from the 
vertex fit.
% (including the $K_s/|Lambda$ secondary vertex if applicable).
%in which
%only the primary charged tracks %and IP information 
%are used.

We perform an unbinned likelihood fit that maximizes the likelihood function, 
%Events are fitted using the extended likelihood function:
$$ L = {e^{-(N_s+N_b)} \over N!}\prod_{i=1}^{N} 
\left[\mathstrut^{\mathstrut}_{\mathstrut}N_s P_s(M_{{\rm bc}_i},\Delta{E}_i)+
N_b P_b(M_{{\rm bc}_i},\Delta{E}_i)\right],$$
to estimate the signal yield in 
5.20 GeV/$c^2 < \mb < 5.29$ GeV/$c^2$ and $-0.1$ GeV $ < \de< 0.2$
GeV;
here $P_s\ (P_b)$ denotes the signal (background) PDF, 
$N$ is the number of events in the fit, and $N_s$ and $N_b$
are fit parameters representing the number of signal and background
events, respectively.

%The PDFs
%are a Gaussian function to represent the signal $\mb$
%and a double Gaussian for $\de$ with parameters determined by
%MC simulation.
For the signal PDF,
we use the product of a Gaussian in $\mb$ and a double Gaussian in $\de$.
We fix
the parameters of these functions to values determined by MC simulation
~\cite{correction}.
The continuum background PDF 
%assumed to be a product function of uncorrelated $\mb$ and $\de$ shapes.
is taken as the product of shapes in
$\mb$ and $\de$, which are assumed to be uncorrelated.
%These shapes are obtained from sideband
%events, with 0.1 GeV $ < \de < 0.2$ GeV for the $\mb$ function and
%with 5.20 GeV/$c^2$ $ < \mb <$ 5.26 GeV/$c^2$ for the $\de$ function.
%They have been cross checked against a continuum MC sample.
We use the parameterization first used by 
the ARGUS collaboration~\cite{Argus}, 
$ f(\mb)\propto \mb\sqrt{1-x^2}
\exp[-\xi (1-x^2)]$,  
%background parametrization first used by the ARGUS collaboration~\cite{Argus} 
to model
the $\mb$ background, with $x$ given by $\mb/E_{\rm beam}$ and $\xi$ as
a fit parameter. %Note that $\mb$ is required to be smaller than $\de$. 
The $\de$ background shape is modeled by a linear function whose slope
is a fit parameter.
% and a first order polynomial 
%straight line 
%for the
%$\de$ background shape. 
%There are possible cross-feeds from
%$\ppkst$ and $\ppksz$ modes to $\ppk$ and $\ppks$
%modes; therefore the cross-feed region ($\de < -0.1$ GeV) 
%is excluded in the fit.
%For $\plpi$ mode, we exclude ($\de < -0.16$ GeV) in the fit.

\begin{figure}[b!]
\centering
%\mbox{\psfig{figure=ppk.phase.1.eps,width=2.5in}}
%\mbox{\psfig{figure=pppi.phase.1.eps,width=2.5in}}
%\mbox{\psfig{figure=ppks.phase.1.eps,width=2.5in}}
%{\bf (a)}\\
\epsfig{file=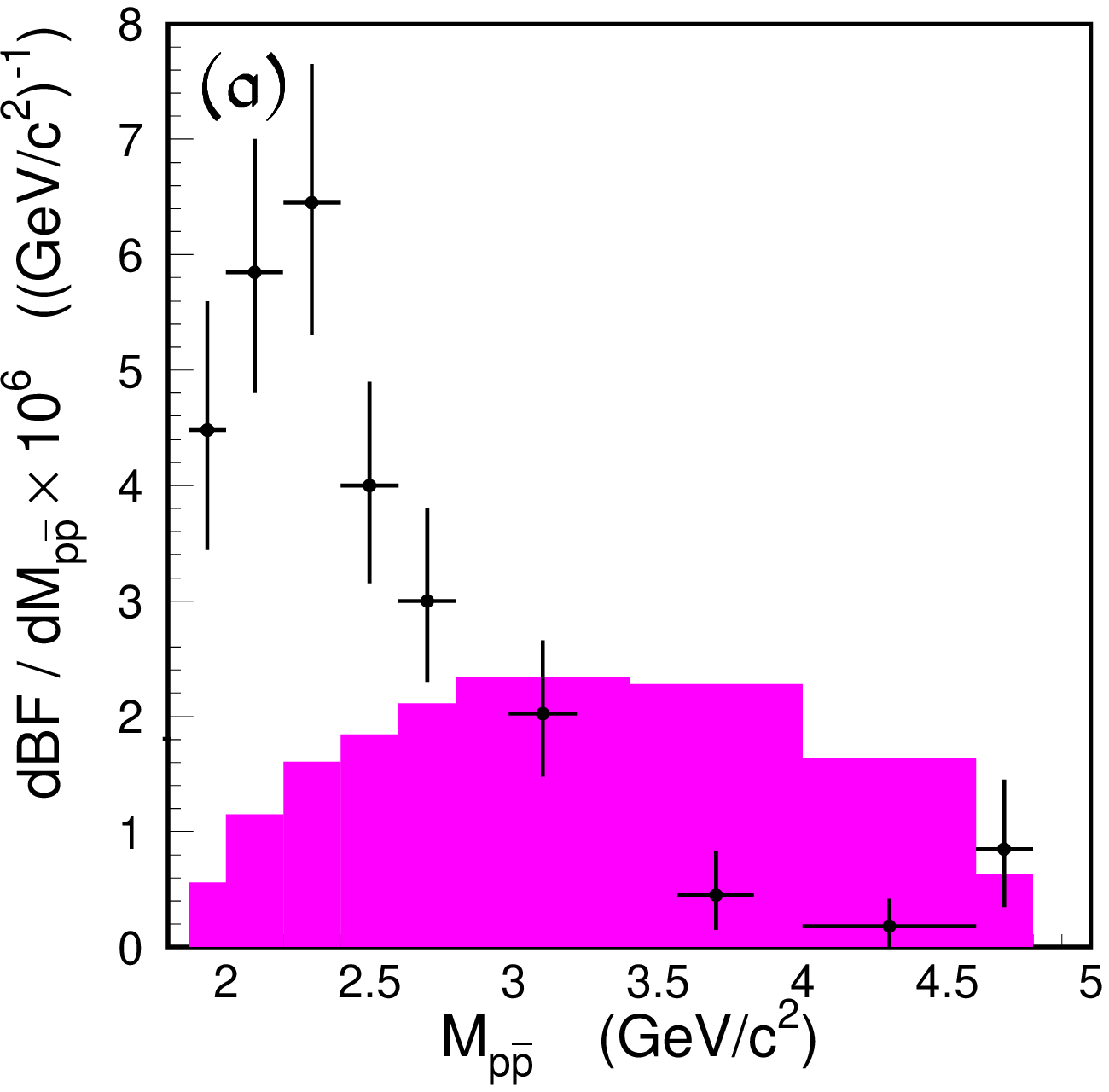,width=7cm}\\
%{\bf (b)}\\
\epsfig{file=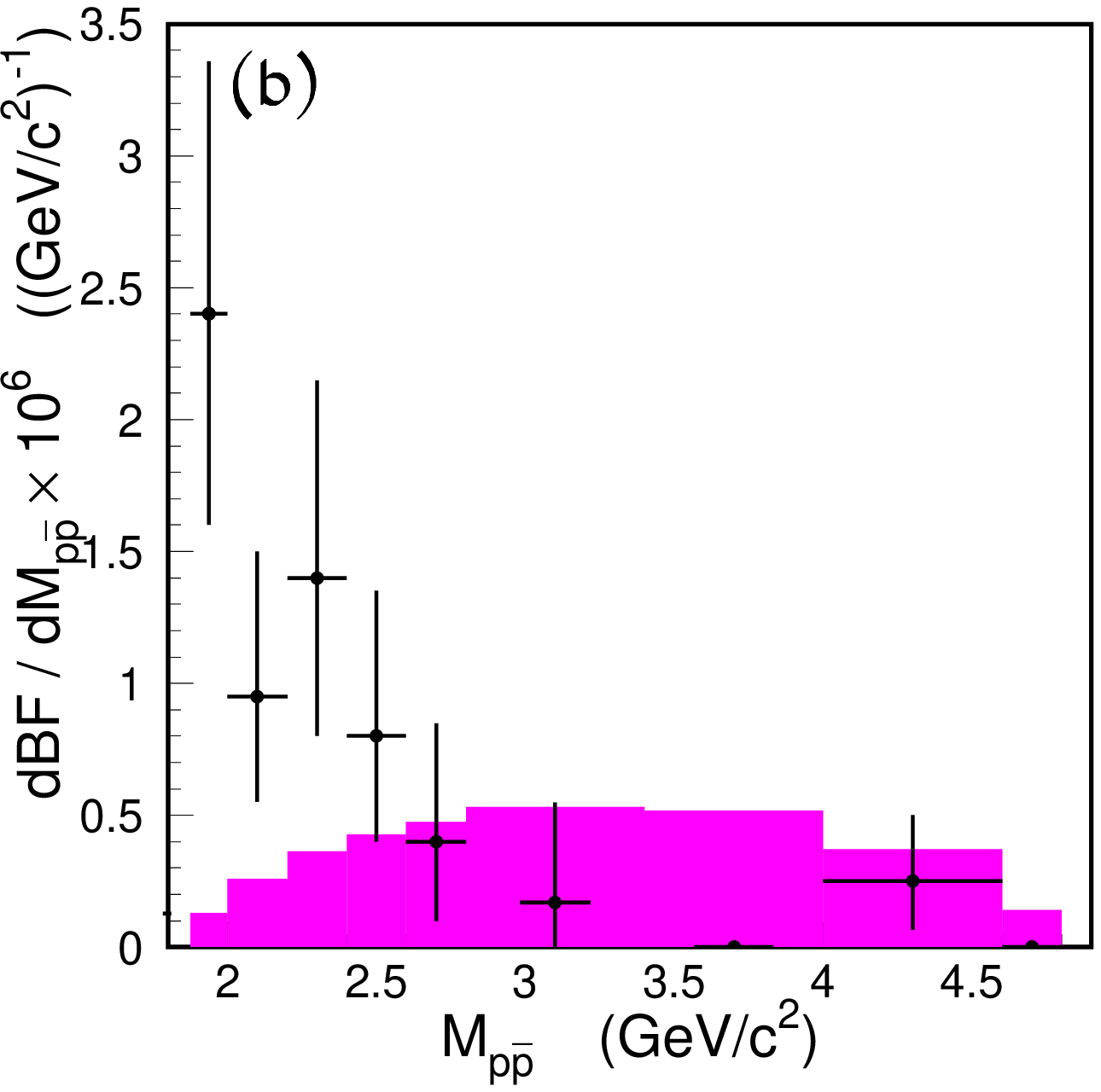,width=7cm}\\
%{\bf (c)}\\
\epsfig{file=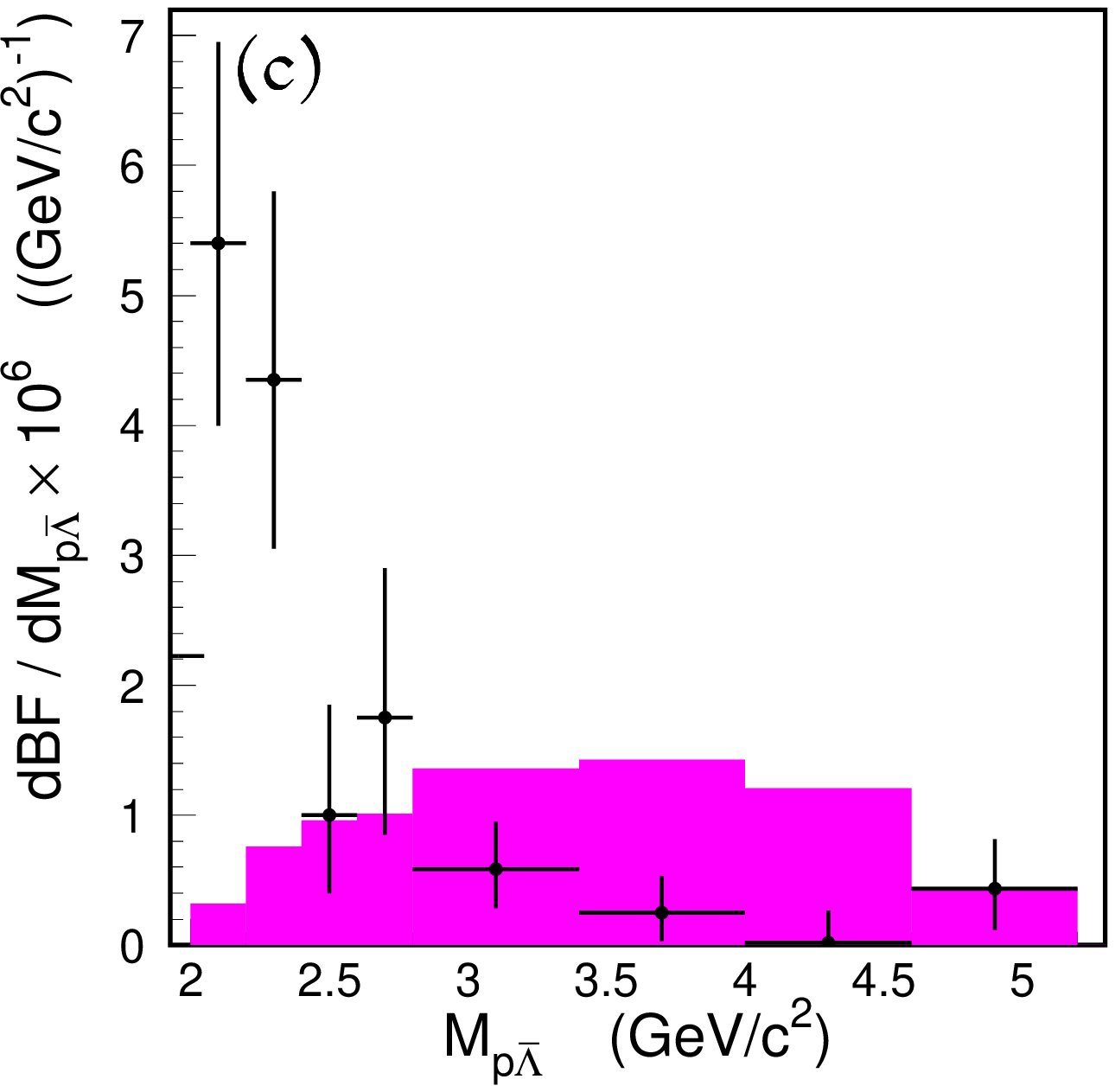,width=7cm}\\

\centering
\caption{Differential branching fraction for %%%divided by the available bin size for
(a) $\ppk$, (b) $\ppks$, and (c) $\plpi$ 
modes %%%in bins
as a function of baryon-antibaryon pair mass. The shaded distribution
shows the expectation from a
phase-space MC simulation with area scaled to the
signal yield. A charmonium veto has been applied in (a) and (b).
%%to modes with $\pp$ pair.
%The insect shows the $p\bar{p}$ mass distribution for the 
%$J/\psi$ region.
}
\label{fg:allphase}
\end{figure}

The differential branching fraction as a function of the 
baryon pair  mass
is shown in Fig.~\ref{fg:allphase}.  
The error bars include the statistical uncertainty from the fit
and the systematic uncertainty. Here, 
the efficiency as a function of baryon pair mass for each signal mode is
determined by MC simulation.
We sum these partial branching fractions to obtain: 
${\mathcal B}(\bp \to \ppk) = (5.30^{+0.45}_{-0.39}
\pm 0.58 )\times 10^{-6}$, ${\mathcal B}(B^0 \to \ppks) =
(1.20^{+0.32}_{-0.22} \pm 0.14) \times 10^{-6}$, 
and ${\mathcal B}(B^0 \to \plpi) = (3.27 ^{+0.62}_{-0.51} \pm 0.39)  
\times 10^{-6}$ which are in good agreement with 
previous measurements~\cite{ppk,plpi,pph}.  
These results also supersede our previous measurements with better accuracy.
%, with the events distributed uniformly in phase space.
%%We generate phase space MC samples and determine the average efficiency
%%in bins of baryon pair mass. 
Note that we have imposed a  charmonium veto for the $\ppk$ and $\ppks$ modes:
the regions  $2.850$ GeV/$c^2  <M_{p \bar{p}}<3.128$ GeV/$c^2$
and $3.315$ GeV/$c^2 <M_{p \bar{p}}<3.735$ GeV/$c^2$
are excluded to remove background from $B$ decay modes
containing an $\eta_c$, $J/\psi$,
$\psi^{\prime}$, $\chi_{c0}$, or $\chi_{c1}$ meson.
%In Fig.~\ref{fg:allphase}, we show
%the differential branching fraction {\it vs} baryon pair  mass.
The width of the low mass enhancement in each distribution of
Fig.~\ref{fg:allphase} depends on the signal mode.
A narrow width is also observed in the newly discovered
%that modes with $\Lambda$'s seem to have a smaller width. The newly discovered
$\bp \to \Lambda \bar{\Lambda} K^+$ decay~\cite{LLK}.
%% decay indeed exhibits the same trend.
%${\mathcal B}(\bz \to \ppkz) = 2{\mathcal B}(\bz \to \ppks)$ is
%assumed. 

Systematic uncertainties %in particle selection
are determined using high statistics control data samples. For proton
identification, we use a  $\Lambda \to p \pi^-$ sample, while for
$K/\pi$ identification we use a $D^{*+} \to D^0\pi^+$,
 $D^0 \to K^-\pi^+$ sample.
%the $\mathcal LR$ selection is studied
%with a
%$B^0 \to D^- \pi^+$, $D^- \to \ks \pi^-$ sample
Tracking efficiency is measured with
fully and partially reconstructed $D^*$ samples.
%$\eta \to \gamma\gamma$ and $\eta \to \pi^+\pi^-\pi^0$ samples.
The uncertainty of $\ks$ reconstruction due to off-IP tracks 
is determined from a $D^- \to \ks\pi^-$
sample. 
The $\Lambda$ and $\ks$ reconstruction efficiencies have the same
uncertainty due to off-IP tracks if the uncertainty of the daughter
proton identification criterion is not taken into account.
The $\mathcal R$ continuum suppression uncertainty is estimated from
 control samples with similar final states, for example,
 $\bp \to J/\psi K^+$  with $J/\psi \to \mu^+\mu^-$.
%$b->c$ control sample
%a $B \to D^0\pi,D^0 \to K\pi$ control sample.
%$\bp \to J/\psi K^+$, $\bz \to J/\psi\ks$, $\bp \to J/\psi\kst$, and
%$\bz \to J/\psi\ksz$
%(with $J/\psi \to \mu^+\mu^-$) control samples.
Based on these studies,
we assign a 1\% error for each track, 3\% for each proton identification,
2\% for each kaon/pion identification, 5\% for $\ks$ and $\Lambda$
off-IP reconstruction
and 6\% for the $\mathcal R$ selection.

A systematic uncertainty of 4\% in the fit yield is determined by varying
the parameters of the signal and background PDFs.  % and is about 4\%. 
%The feed down effect from
%$\ppk$ to $\pppi$ makes the total fit error of the $\pppi$ mode at
%the 5\% level. 
The MC statistical
uncertainty %and binning of the baryon pair mass 
contributes a 2\%
error in the branching fraction determination. The error on the
number of $B\bar{B}$ pairs is 0.5\%, where we
assume  that the branching fractions of $\Upsilon({\rm 4S})$ 
to neutral and charged $B\bar{B}$ pairs are equal. 
%~\cite{Bellenote}

We first sum the correlated errors linearly and then combine them with the
uncorrelated ones in quadrature. The total systematic
uncertainties are 11\%, 12\%, and 12\% for
the $\ppk$, $\ppks$, and $\plpi$ modes,
respectively.

\begin{figure}[p]
%\vskip -2in
\centering
%%%{\bf  (a)}\\
\epsfig{file=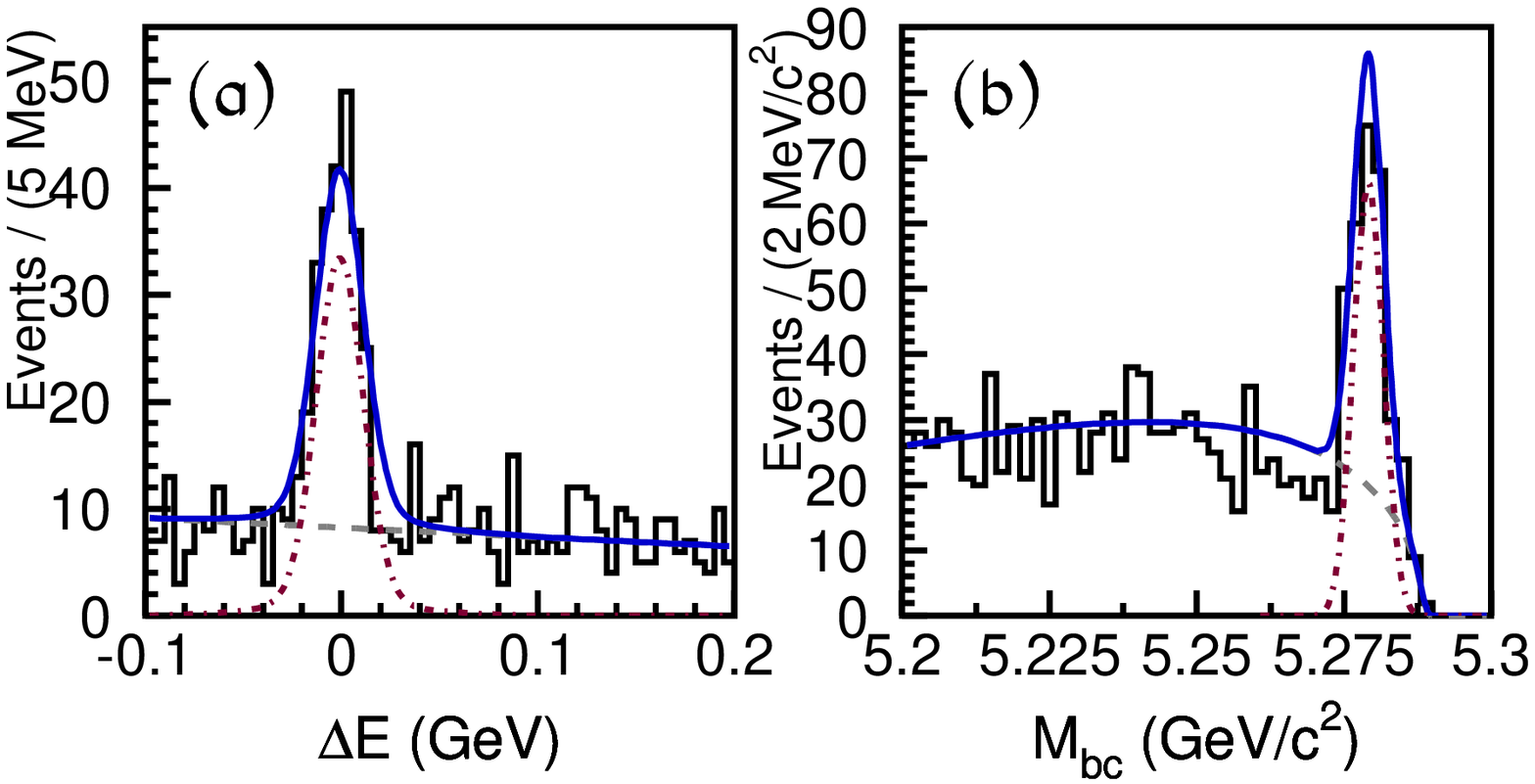,width=10cm}\\

%%%{\bf  (b)}\\
\epsfig{file=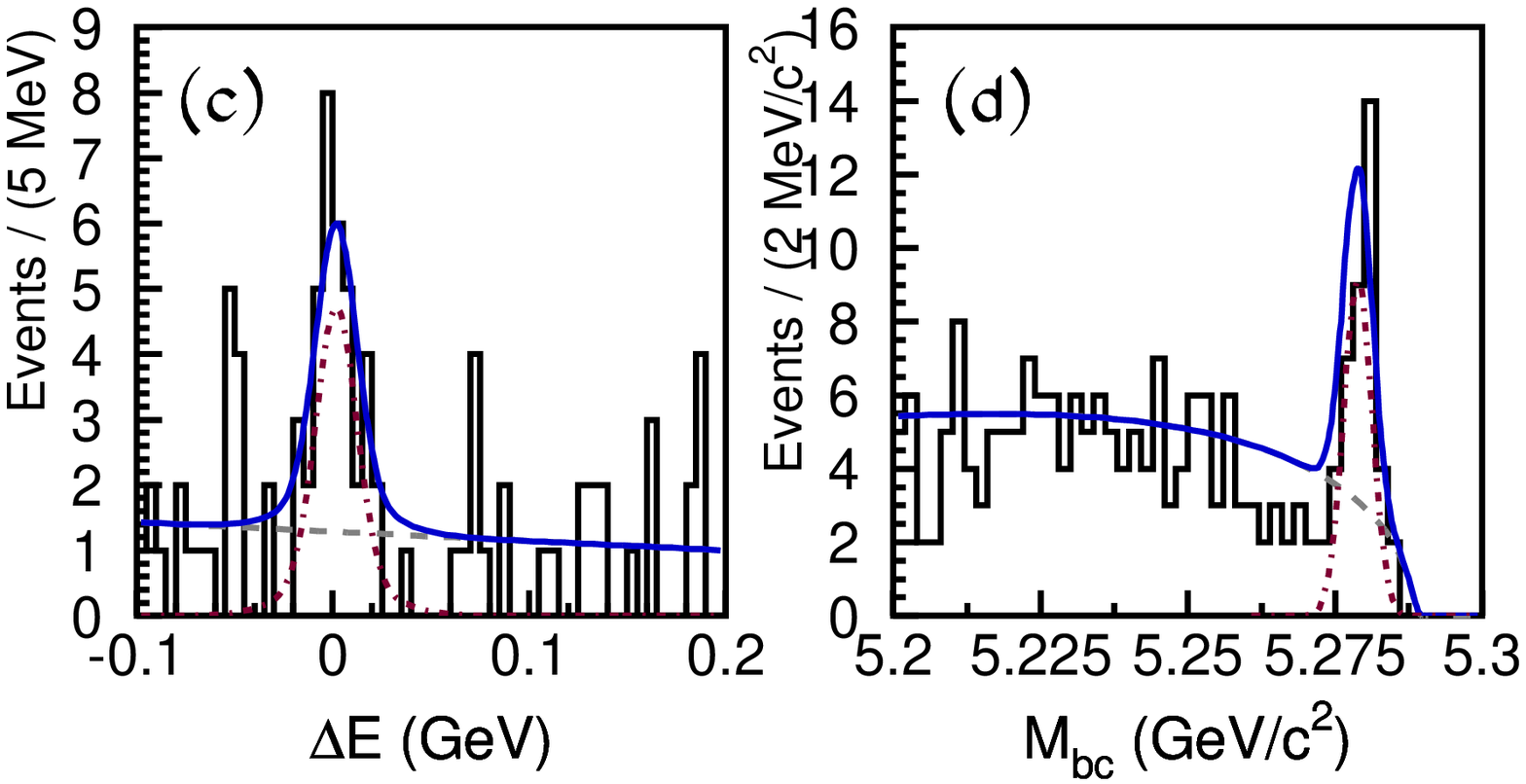,width=10cm}\\

%%%{\bf (c)}\\
\epsfig{file=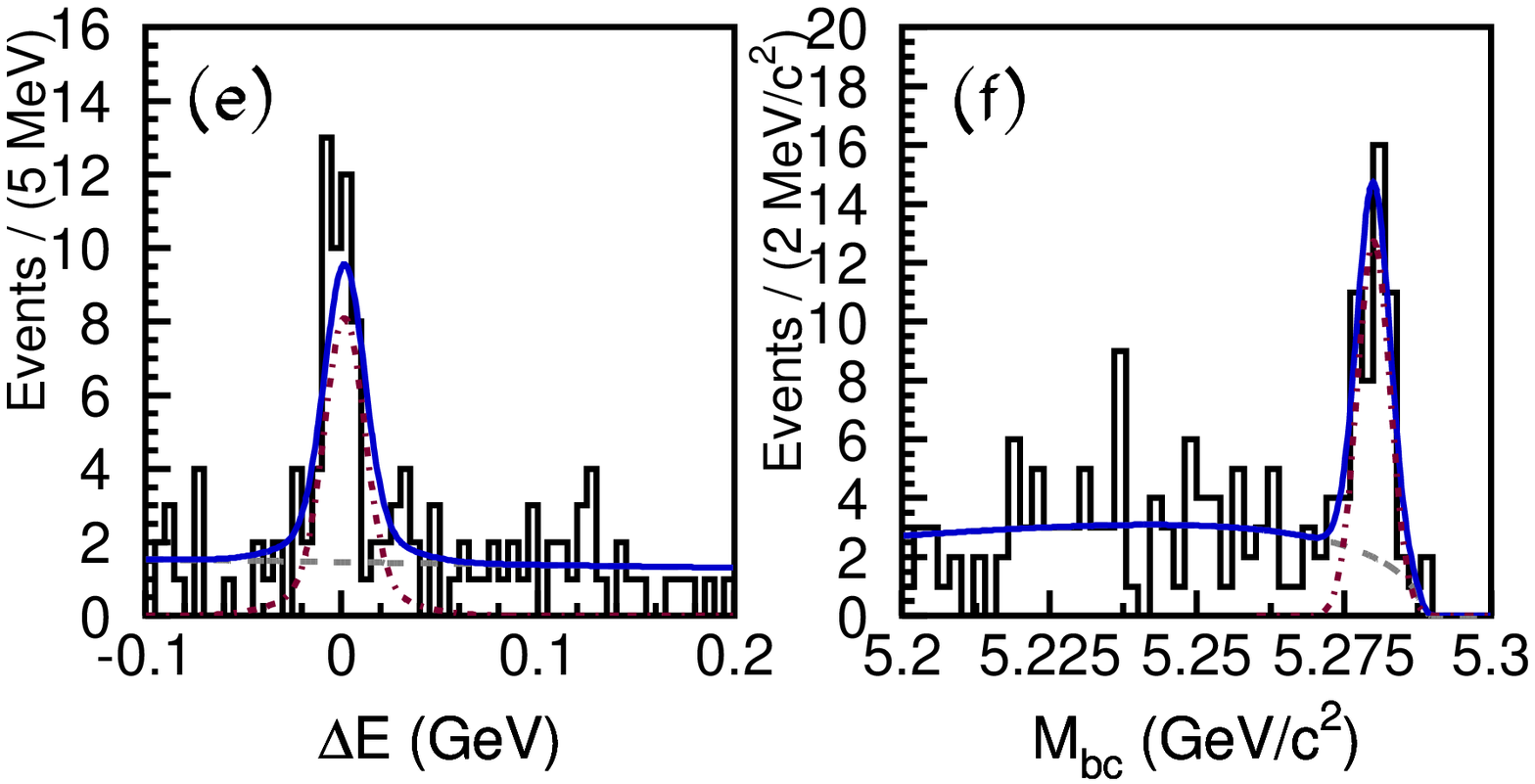,width=10cm}\\

%\epsfig{file=mergembppksz.1.eps,width=5cm,height=4.3cm} 
%\hskip 1cm 
%\epsfig{file=mergedeppksz.1.eps,width=5cm,height=4.3cm} 
\centering
\caption{ Distributions of $\de$ and $\mb$, respectively, for
(a) and (b) $\ppk$, (c) and (d) $\ppks$, and (e) and (f) $\plpi$ 
modes with baryon-antibaryon pair mass less than  
2.85 GeV/$c^2$. 
The solid, dotted and dashed lines represent the combined fit
result, fitted signal and fitted background, respectively.
}

\label{fg:mergembde}
\end{figure}

%Based on the above measurements, 
%showed an enhancement at
%low baryon-antibaryon  mass, we first focus on the near
%threshold region by 
We require the  mass of the baryon pair to be less than 
%$\mpp <$ 
2.85 GeV/$c^2$ for the study of the threshold enhancement effect.
%for making sure it
%is below charmonium threshold.
%Since these decays do not follow three-body phase space decay,
%one can focus on
%the near threshold region.
The $\mb$ distributions (with $|\de|<$ 0.05
GeV), and the $\de$ distributions (with $\mb >$ 5.27 GeV/$c^2$)
for the $\ppk$, $\ppks$ and $\plpi$ modes are
shown in Fig.~\ref{fg:mergembde}. 
The projections of the fit results are shown in 
Fig.~\ref{fg:mergembde} by solid curves. 
%Since there is no clear evidence for the $\ppksz$
%mode, we plot the 90\% confidence level upper limit in this figure.
The $B$ yields are
%217 $^{+17}_{-17}$,
$217 \pm 17$,
28.6 $^{+6.5}_{-5.8}$,
and 48.8 $^{+8.2}_{-7.5}$ 
%with
%significances of $15.3$, $6.7$, $5.1$, and $6.0$ standard
%deviations 
for the $\ppk$, $\ppks$, and $\plpi$ modes, respectively. 
The measured 
branching fractions by summing the partial branching fractions 
in mass bins below 2.85 GeV/$c^2$ are
${\mathcal B}(\bp \to \ppk) = (4.59^{+0.38}_{-0.34}
\pm 0.50 )\times 10^{-6}$, ${\mathcal B}(B^0 \to \ppks) =
(1.04^{+0.26}_{-0.19} \pm 0.12) \times 10^{-6}$, 
and ${\mathcal B}(B^0 \to \plpi) = (2.62 ^{+0.44}_{-0.40} \pm 0.31)  
\times 10^{-6}$. 
%The 
%significance is defined as $\sqrt{-2 {\rm ln}(L_0/L_{max})}$~\cite{PDG}, 
%where $L_0$ and
%$L_{max}$ denote the likelihood with signal yield fixed at
%zero and at the fitted value, respectively. 
%by the maximum value of the likelihood function
%and the likelihood value with yield fixed at zero~\cite{PDG}.
%The yield for $\ppksz$ is less significant.
%The non-resonant $\ks \pi^+$ component of the 
%$\ppkst$ mode is included in the
%systematic error estimation as described later.

We study the proton angular distribution of the baryon-antibaryon pair system
in its helicity 
frame. The angle $\theta_p$ is defined as the
angle between the proton direction and the meson direction in the 
baryon-antibaryon pair rest frame.  
Note that the angle is determined by $\bar{p}$ and $K^+$ 
(or $p$ and $K^-$) 
in the $\ppk$ mode for definiteness. 
Fig.~\ref{fg:thetap}(a)-(c) shows the
branching fractions as a function of $\cos \theta_p$. 
%The error bars
%include the statistical uncertainty from the fit
%and the systematic uncertainty. 
We define the angular asymmetry as $A = {
{N_+ - N_-}\over
{N_+ + N_-}}$, where $N_+$ and
$N_-$
stand for the efficiency corrected $B$ yield with $\cos\theta_p > 0$ and
 $\cos\theta_p < 0$, respectively. The angular asymmetry is determined to be
$0.59^{+0.08}_{-0.07}$ for the $\ppk$ mode.
The asymmetry of the distribution indicates that the fragmentation
picture is favored. 
%since a symmetric distribution is 
%preferred for the gluonic picture. 
Antiprotons are emitted along
the $K^+$ direction most of the time, which can be explained by 
a parent $\bar{b} \to \bar{s}$ penguin transition followed by
$\bar{s} u$ fragmentation into the final state as shown in Fig.~\ref{fg:feyn}. 
The energetic $\bar{s}$ quark picks up the $u$ quark from a
$u\bar{u}$ pair in vacuum and the remaining $\bar{u}$ quark then
drags a $\bar{u}\bar{d}$ diquark out of vacuum. 
This simple picture can describe the
$\bar{p}-K^+$ angular correlation. The spectator $u$ quark and leftover
$u d$ diquark form an proton.     

The $\cos \theta_p$ distribution of the $\ppks$ mode 
can not support nor refute this fragmentation interpretation because
of low statistics and no flavor information. 
%although it seems to be
%peaked toward $\pm 1$
%%a more or less symmetric pattern with a peaking tendency towards $\pm 1$ 
%since the flavor information is not applied in this case.
%%not considered here. 
The distribution for the $\plpi$ mode 
% cannot be explained by the fragmentation process. It 
is quite flat (i.e. in favor of the gluonic picture),  
in contrast to that of the
%%%does not have a visible correlation like
$\ppk$ mode, although both presumably share a common origin in the
$\bar{b} \to \bar{s}$ transition. 
%%%In order to make a fair comparison, we re-draw
In fact, this parentage suggests that it would be useful to examine
the proton angular distribution 
relative to the $\bar{\Lambda}$ direction in
the $p \pi^-$ rest frame.
We remove the $M_{p\bar{\Lambda}} < 2.85$ GeV/$c^2$ constraint 
in order to check the full angular region;
the result is shown in Fig.~\ref{fg:thetap}(d). It is evident 
that the fragmentation interpretation is supported: 
%%%fragmentation picture is still hold.
the proton tends to emerge parallel to the $\bar{\Lambda}$ baryon.
%%%are generated in parallel for most of the time.

As a cross check, the distribution of $\cos \theta_p$
for background events in the $\ppk$ sample
is shown in Fig.~\ref{fg:thetap}(e). Similar distributions are obtained 
for the backgrounds of the $\ppks$ and $\plpi$ modes. The background has
a $1+\alpha\cos^2 \theta_p $ distribution, which can be explained  as
arising from the random combination of two high momentum particles from the
$q \bar{q}$ jets.  The continuum MC simulation also confirms this feature.
%The fragmentation signature is not seen in
%%%Another check is done with the
%the $B^+ \to J/\psi K^+$ mode, where the $J/\psi$ meson decays to a
%$\pp$ pair.
%For $J/\psi$ candidates with  mass in the range
%$3.07$ GeV/$c^2 < \mpp < 3.11$ GeV/$c^2$,
%the distribution of $\cos \theta_p$ is flat,
%as shown in Fig.~\ref{fg:thetap}(f).
%%%There is no such fragmentation signature for $\pp$ resonance decay.

\begin{figure}[b!]
\centering
%\mbox{\psfig{figure=ppk.phase.1.eps,width=2.5in}}
%\mbox{\psfig{figure=pppi.phase.1.eps,width=2.5in}}
%\mbox{\psfig{figure=ppks.phase.1.eps,width=2.5in}}
%\centering
\epsfig{file=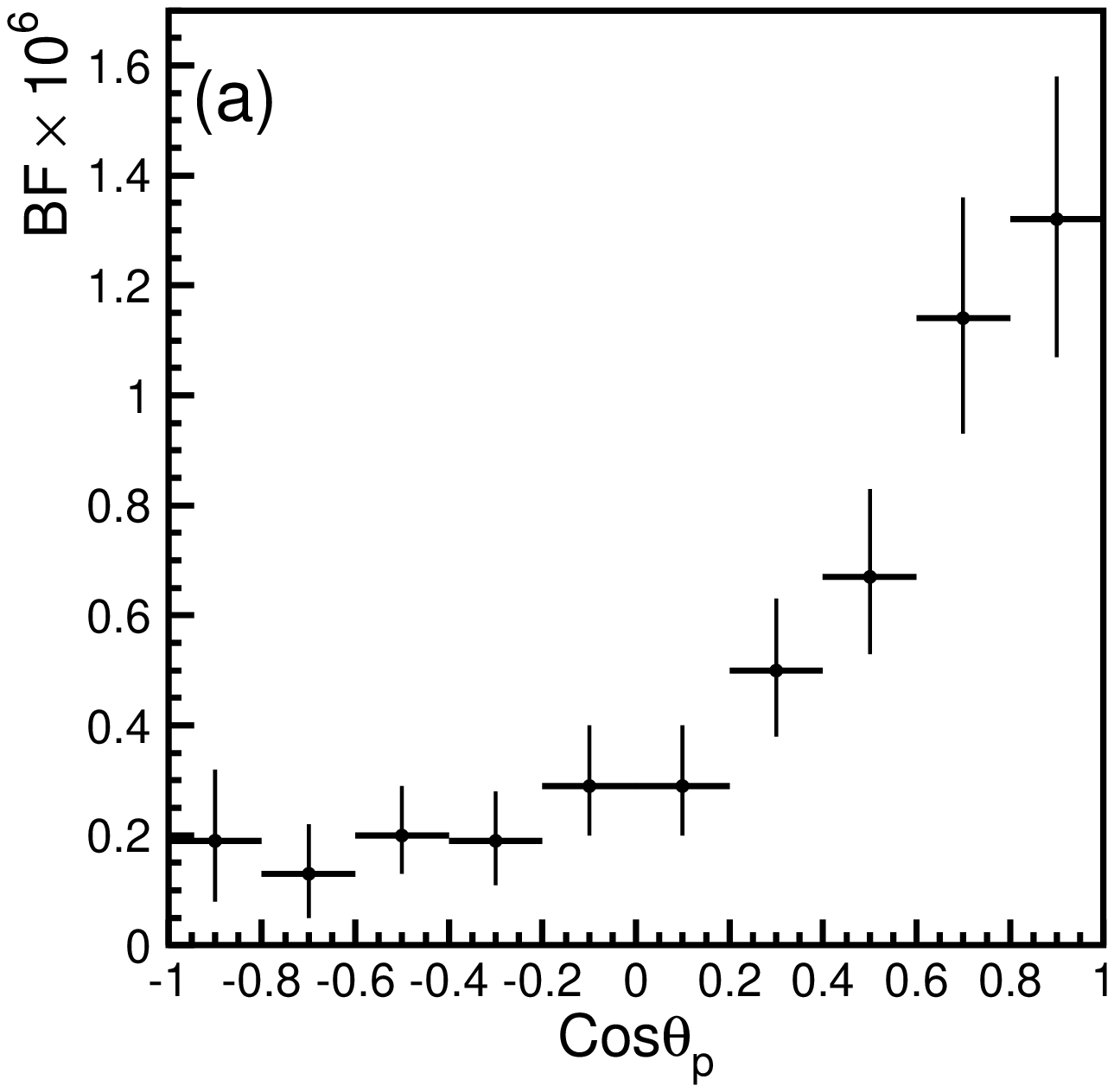,width=5cm}\enspace%
\epsfig{file=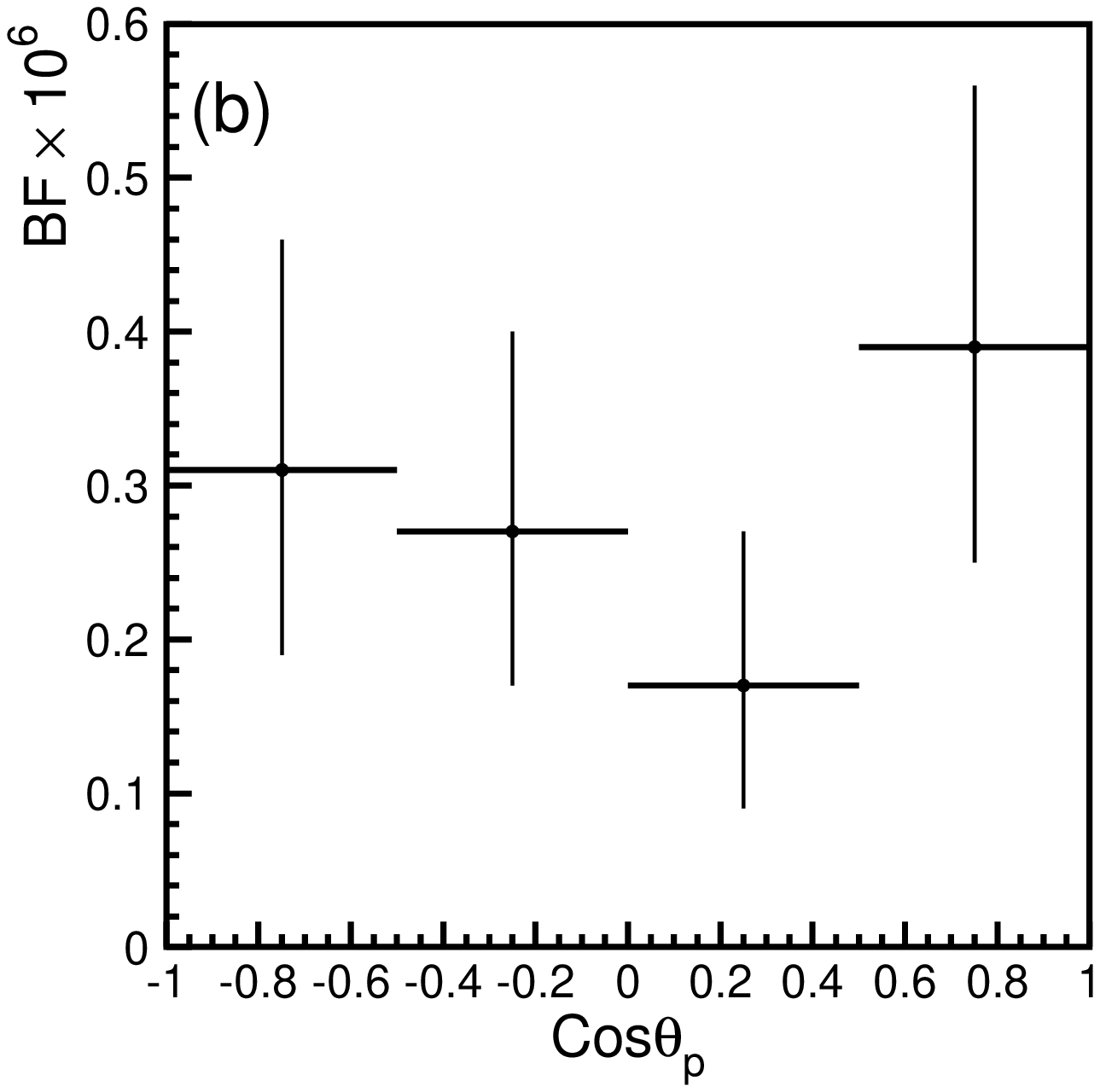,width=5cm}\enspace%
\epsfig{file=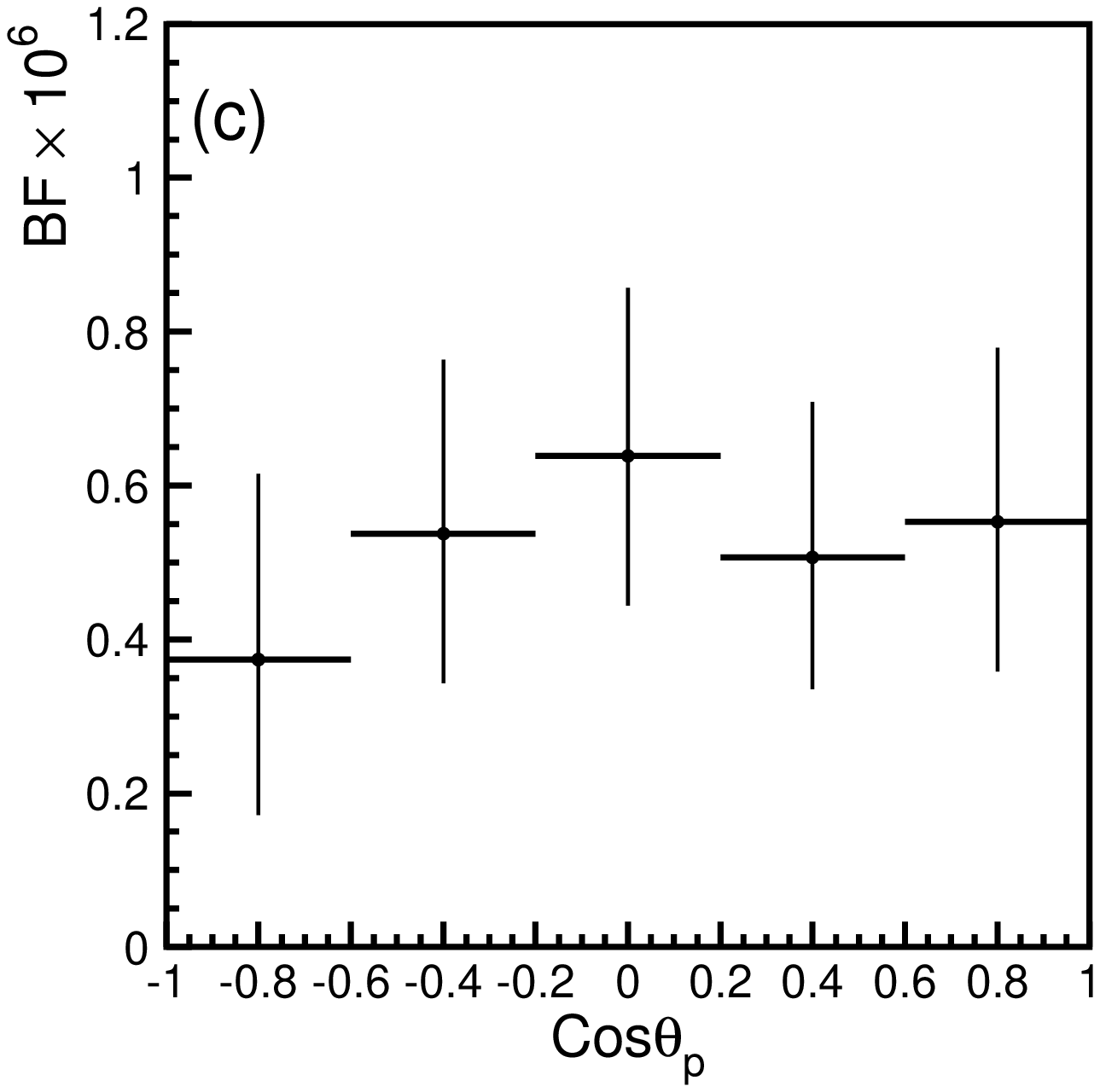,width=5cm}\\
\epsfig{file=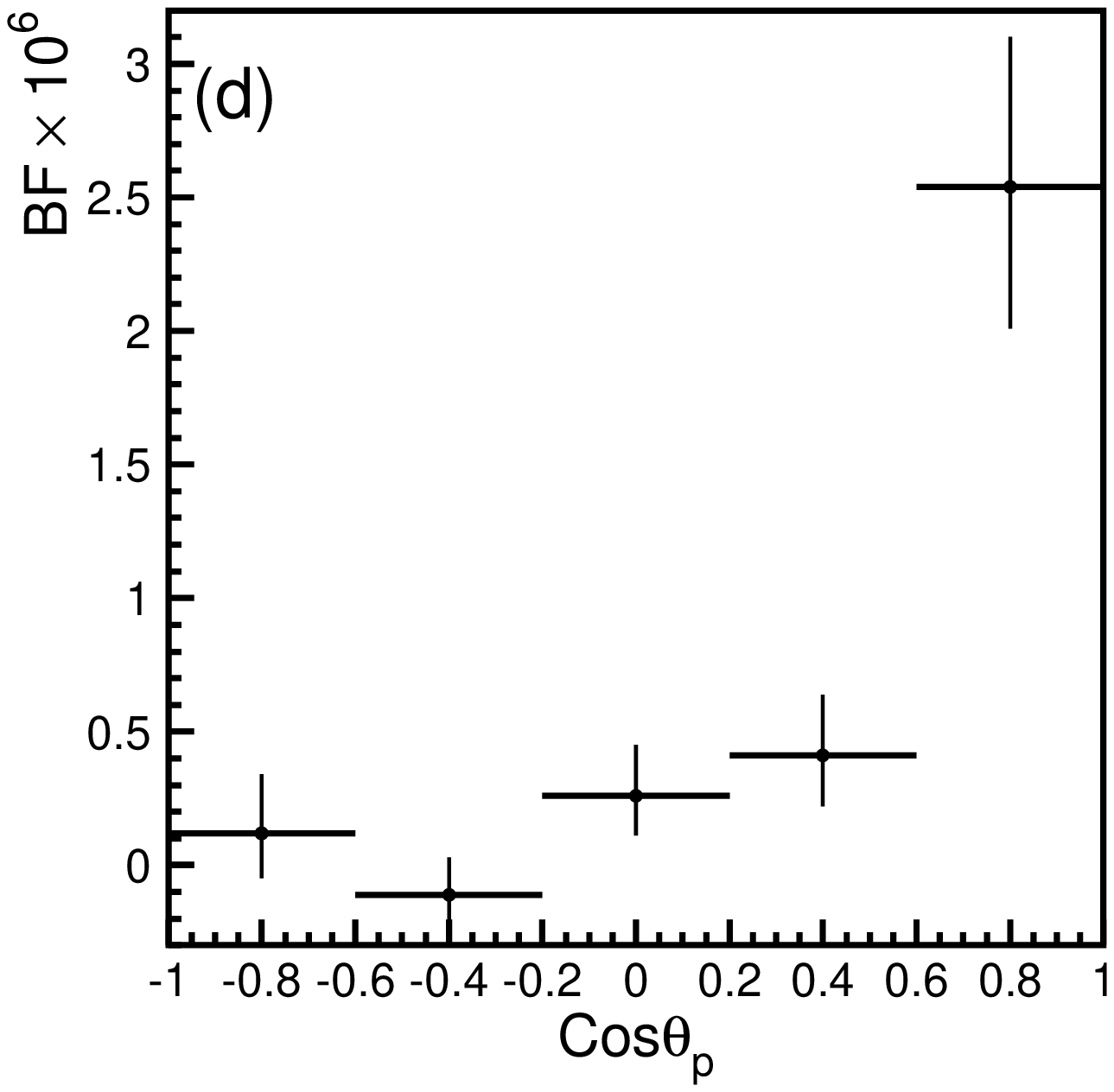,width=5cm}\enspace%
\epsfig{file=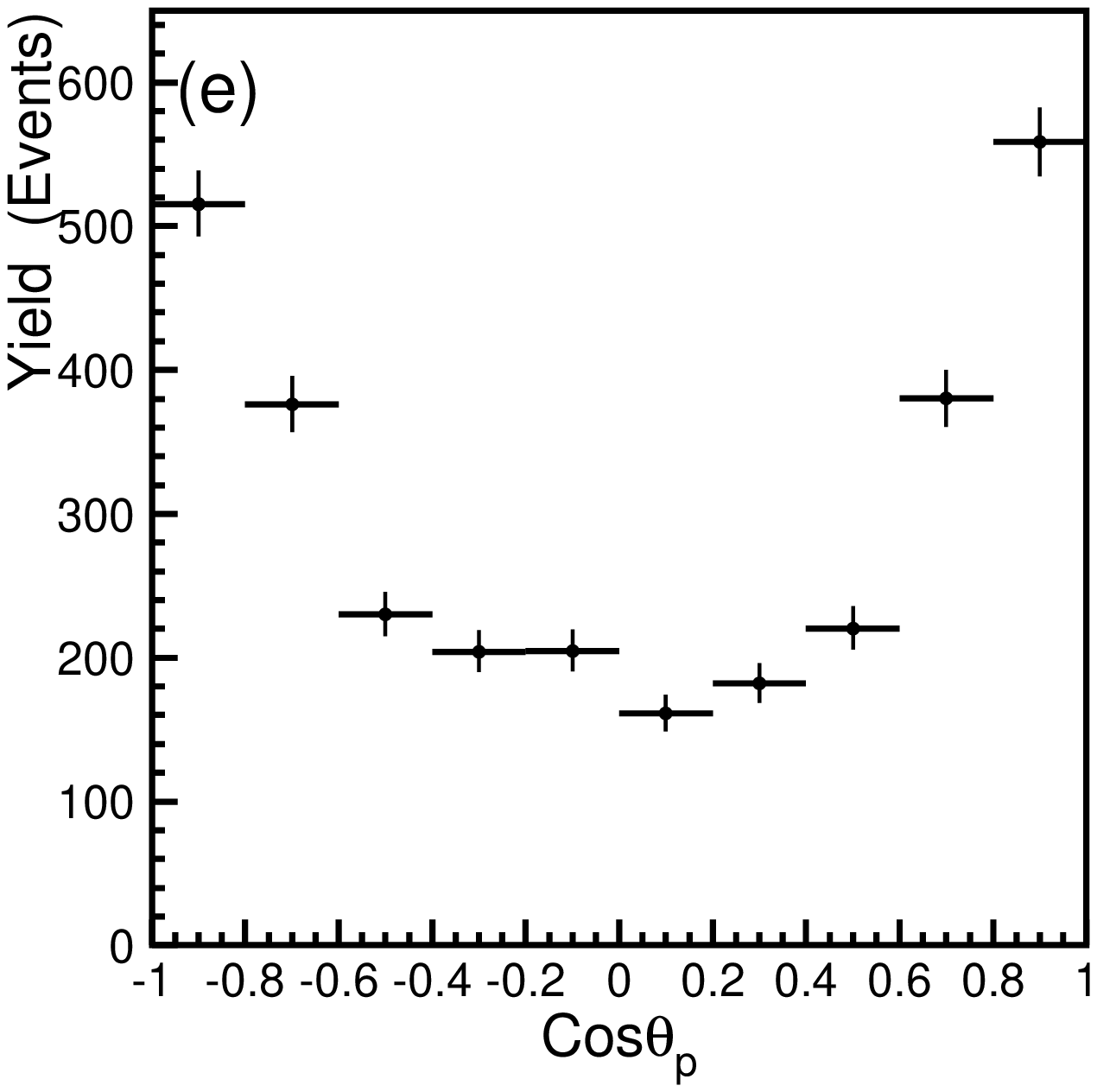,width=5cm}\enspace%
\centering
\caption{Branching fraction {\it vs.} $\cos \theta_p$
 in the baryon-antibaryon pair system for
(a) $\ppk$, (b) $\ppks$, and (c) $\plpi$ modes. (d) The proton angular
distribution of the $p \pi^-$ system %%%in the $\plpi$ mode 
against the $\bar{\Lambda}$ direction %%%is shown in (d). 
in the $\plpi$ mode.
%%%The distribution
(e) $\ppk$ background yield {\it vs.} $\cos \theta_p$. 
%(f) Comparison with the
%$J/\psi$ mass region. %%% is shown in (f).  
}
\label{fg:thetap}
\end{figure}

\begin{figure}[p]
%\vskip -2in
\centering
\epsfig{file=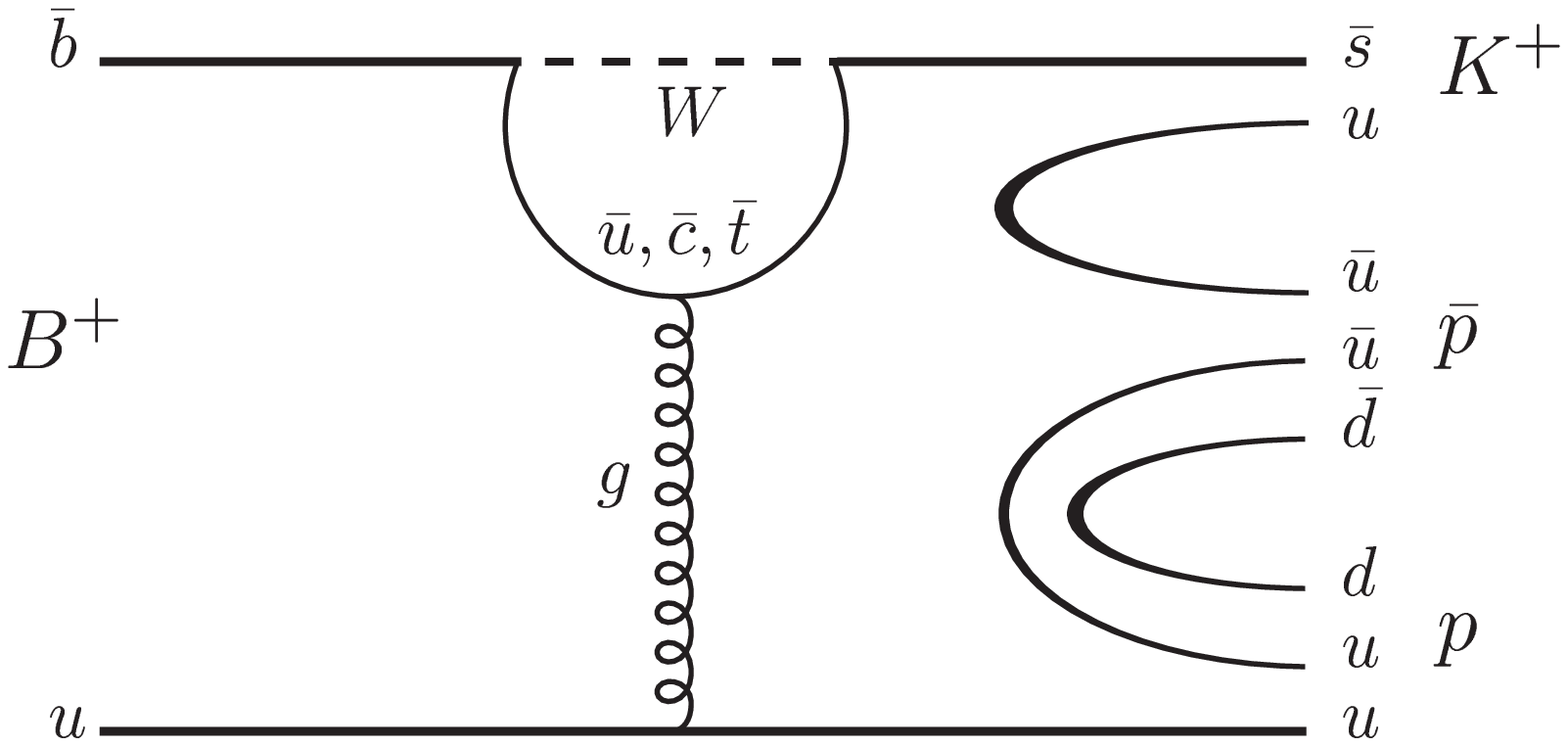,width=10cm}\\
\centering
\caption{
Simple quark diagram for an illustration of the $B^+ \to \ppk$ decay.  
}
\label{fg:feyn}
\end{figure}

The newly observed narrow pentaquark state, $\Theta^+$~\cite{penta}, can
decay into $p \ks$. We perform a search in $B^0 \to \ppks$
by requiring   $1.53$ GeV/$c^2 < M_{p\ks} < 1.55 $ GeV/$c^2$.
The $\mb$ and $\de$ projection plots in Fig.~\ref{fg:penta} show
no evidence for a pentaquark signal. Since there are few events in the fit
 window, we fix the
background shapes from sideband data.
We use the fit results to estimate the expected background ($0.42 \pm 0.13$)
and compare this  with the observed one event 
in the signal region
to set an upper limit on the signal
yield of 3.9 events at the 90\% confidence level~\cite{Gary,Conrad}. The
systematic uncertainty is included in this limit. 
%The determined upper limit yield on top of the background
%is shown as solid line
%in Fig.~\ref{fg:mergembde}(e).
The related upper limit on the product of branching fractions is 
${\mathcal B}(B^0 \to {\Theta^+}\bar{p})\times {\mathcal B}({\Theta^+
}\to p\ks) < 2.3 \times 10^{-7}$.
We also perform a search for $\Theta^{++}$, 
which can decay to $p K^+$ in the mode $B^+ \to \ppk$~\cite{pentappk}.
Because there are only theoretical conjectures for the existence of 
such a state, we examine the wider mass region of
$1.6$ GeV/$c^2 < M_{p K^+} < 1.8 $ GeV/$c^2$.  We
find no evidence for signal. Assuming this state is narrow and 
centered near 1.71 GeV/$c^2$, the
upper limit on the yield is 3.3 events at the 90\% confidence level. The
corresponding upper limit product of branching fractions is
${\mathcal B}(\bp \to {\Theta^{++}}\bar{p})\times {\mathcal B}({\Theta^{++}
}\to p K^+) < 9.1 \times 10^{-8}$ at the 90\% confidence level.

\begin{figure}[p]
%\vskip -2in
\centering
\epsfig{file=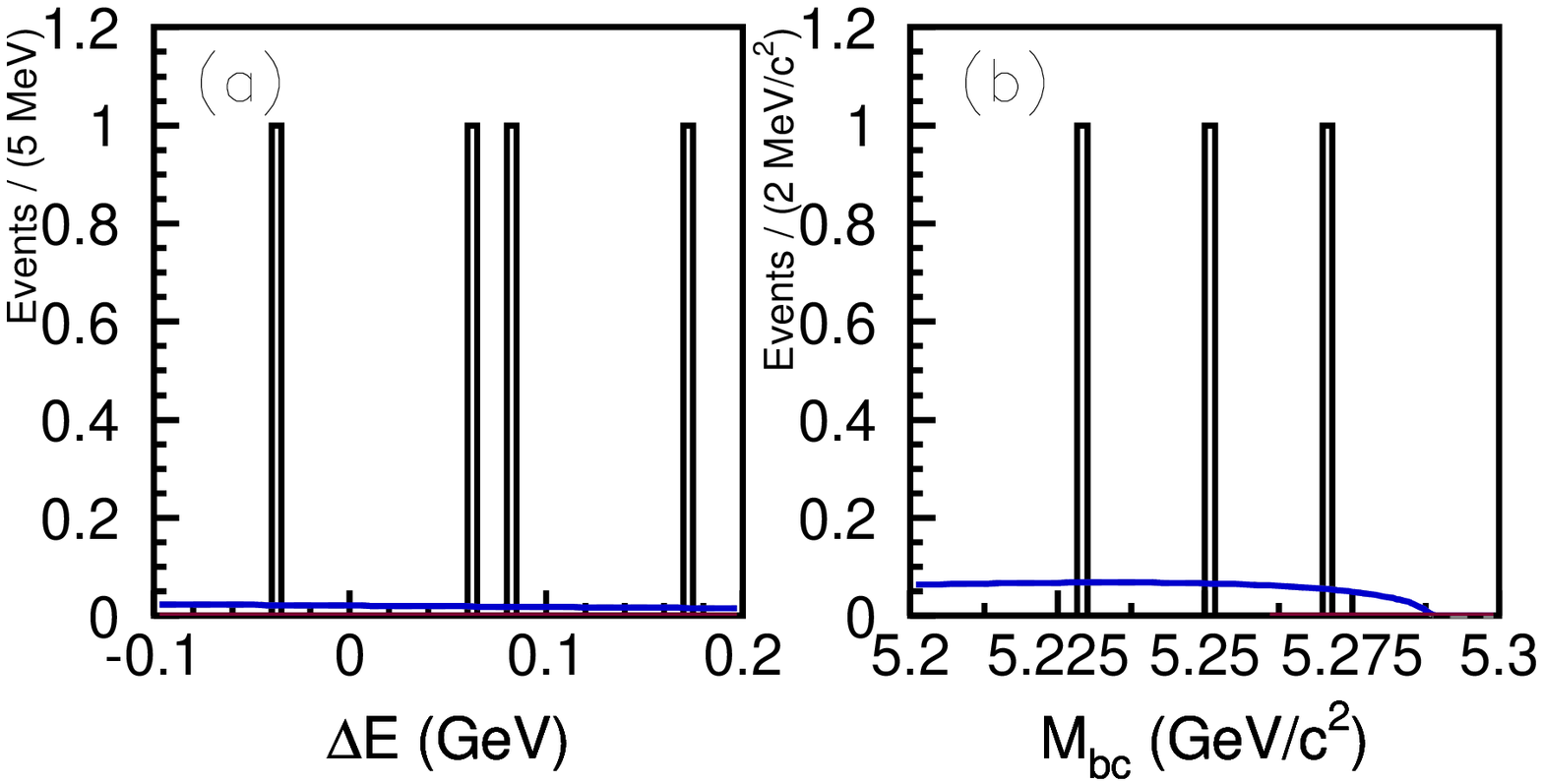,width=10cm}\\
\centering
\caption{Distributions of $\mb$ and $\de$ for the $\ppks$ mode
with $1.53$ GeV/$c^2 < M_{p\ks} < 1.55 $ GeV/$c^2$ in the region where the
$\Theta^+$ pentaquark is expected.
The curves represent the fit
projections. 
% except in (e) where the
%curve is the 90\% confidence level signal yield superimposed 
%on the background fit.
}

\label{fg:penta}
\end{figure}

One theoretical
conjecture~\cite{glueball} suggests that a possible 
glueball resonance $\mathcal{G}$ (e.g., $f_J(2220)$~\cite{PDG}) 
with mass
near 2.3
GeV/$c^2$ may contribute to the $\mpp$ threshold
peaking behavior for the $\ppk$ mode.  
Since the $\mpp$ mass
resolution is about 10 MeV/$c^2$, we scan through the
%% 20 MeV/$c^2$ bin size scanning
$2.2$ GeV/$c^2 < \mpp < 2.4 $ GeV/$c^2$ mass region
with a 20 MeV/$c^2$ wide window. The largest upper 
limit on the yield is found to be 18.9 at 2.21 GeV/$c^2$. 
We use this data set to set an upper limit on the product of branching
fractions of
${\mathcal B}(\bp \to \mathcal{G} K^+)\times {\mathcal B}( \mathcal{G} \to 
\pp) < 4.1 \times 10^{-7}$ at the 90\%
confidence level for a narrow glueball state with  mass
in the 2.2 -- 2.4 GeV/$c^2$ range. The theoretical expectation is
around $1 \times 10^{-6}$.

In summary, using 152 $ \times 10^6 B\bar{B}$ events, we measure the
 mass and the angular distributions of the baryon-antibaryon pair
system near threshold for the $\ppk$, $\ppks$ and $\plpi$  
baryonic $B$ decay modes. A quark fragmentation
interpretation is supported, while a gluonic resonant state 
picture is disfavored. 
%It plays an important role  
%for the threshold enhancement effect. 
%It will be interesting to compare
%with other observed decay modes (such as $\pppi$ and $\llk$) to learn
% more about the underlying dynamics. 
%We need more observations to understand the underlying dynamics.
Searches for a $B$ meson decaying into the $\Theta^+$ pentaquark or a
glueball in the above related modes give null results.  

%\section*{Acknowledgments}
We thank the KEKB group for the excellent operation of the
accelerator, the KEK cryogenics group for the efficient
operation of the solenoid, and the KEK computer group and
the National Institute of Informatics for valuable computing
and Super-SINET network support. We acknowledge support from
the Ministry of Education, Culture, Sports, Science, and
Technology of Japan and the Japan Society for the Promotion
of Science; the Australian Research Council and the
Australian Department of Education, Science and Training;
the National Science Foundation of China under contract
No.~10175071; the Department of Science and Technology of
India; the BK21 program of the Ministry of Education of
Korea and the CHEP SRC program of the Korea Science and
Engineering Foundation; the Polish State Committee for
Scientific Research under contract No.~2P03B 01324; the
Ministry of Science and Technology of the Russian
Federation; the Ministry of Higher Education, Science and
Technology of the Republic of Slovenia;
the Swiss National Science Foundation; the National Science
Council and the Ministry of Education of Taiwan; and
the U.S. Department of Energy.


\begin{thebibliography}{99}
\bibitem{conjugate} {Throughout this report,
inclusion of charge conjugate mode is always implied unless
otherwise stated.}
\bibitem{correction}{ There are small corrections applied to these
parameters based on the 
study of the difference between data and MC in $B \to D\pi$ decays.}
\end{thebibliography}
\end{document}